\documentclass[sigconf]{acmart}

\renewcommand\footnotetextcopyrightpermission[1]{}

\newcommand{\black}[1]{\textcolor{black}{#1}}

\usepackage{soul}
\newcommand{\drop}[1]{\textcolor{red}{#1}}
\renewcommand{\drop}[1]{}
\definecolor{cadmiumgreen}{rgb}{0.0, 0.42, 0.24}
\usepackage{multirow}
\usepackage{enumitem}

\usepackage{enumitem}
\usepackage{graphicx}
\usepackage{booktabs} 
\usepackage{pifont}
\usepackage{mathtools}
\def\smallerspacecaption{\vspace{-2mm}}

\usepackage[ruled, vlined, norelsize]{algorithm2e} 
\SetKwInput{KwInput}{Input}
\SetKwInput{KwOutput}{Output}
\SetKwInput{KwInit}{Initialization}
\SetKwInput{Kwprocedure}{Procedure}
\usepackage{lipsum}
\usepackage{dblfloatfix}

\usepackage{bm}

\usepackage{scalerel}[2016/12/29]

\definecolor{green}{HTML}{2a7e6b}

\newcommand{\ssub}[2]{{#1}_{\scaleobj{0.8}{#2}}}

\usepackage{blindtext,graphicx}

\usepackage{bm}
\usepackage{circledsteps}
\pgfkeys{/csteps/inner color=white}
\pgfkeys{/csteps/outer color=none}
\pgfkeys{/csteps/fill color=black}

\usepackage{algorithmic}
\usepackage{textcomp}
\AtBeginDocument{%
  \providecommand\BibTeX{{%
    \normalfont B\kern-0.5em{\scshape i\kern-0.25em b}\kern-0.8em\TeX}}}

\def\smallerspacecaption{\vspace{-2mm}}
\usepackage{setspace}

\usepackage{multirow}
\usepackage{graphicx}
\usepackage{amsmath,amsfonts,bm}

\newcommand{\func}[2][]{
    {\text{\fontfamily{lmtt}\selectfont #1}\left(#2\right)}}
\def\mA{{\bm{A}}}
\def\ma{{\bm{a}}}
\def\mz{{\bm{z}}}

\def\mx{{\bm{x}}}

\def\mD{{\bm{D}}}

\def\mI{{\bm{I}}}

\def\mW{{\bm{W}}}
\def\mX{{\bm{X}}}

\def\mZ{{\bm{Z}}}

\DeclareMathAlphabet{\mathsfit}{\encodingdefault}{\sfdefault}{m}{sl}
\SetMathAlphabet{\mathsfit}{bold}{\encodingdefault}{\sfdefault}{bx}{n}

\def\gG{{\mathcal{G}}}

\def\gN{{\mathcal{N}}}

\newcommand{\R}{\mathbb{R}}

\begin{document}

\title{Embracing Graph Neural Networks for Hardware Security (Invited Paper)}
\author{Lilas~Alrahis$^\ddag$, Satwik~Patnaik$^\dag$, 
Muhammad~Shafique$^\ddag$, and Ozgur~Sinanoglu$^\ddag$\\[1ex]
$^\ddag$Division of Engineering, New York University Abu Dhabi, UAE\\
$^\dag$Electrical \& Computer Engineering, Texas A\&M University, College Station, Texas, USA\\
\normalsize{\{lma387, muhammad.shafique, ozgursin\}@nyu.edu},
\normalsize{satwik.patnaik@tamu.edu}
}

\renewcommand{\shortauthors}{Alrahis et al.}

\begin{abstract}
Graph neural networks (GNNs) have attracted increasing attention due to their superior performance in deep learning on graph-structured data. GNNs have succeeded across various domains such as social networks, chemistry, and electronic design automation (EDA). 
Electronic circuits have a long history of being represented as graphs, and to no surprise, GNNs have demonstrated state-of-the-art performance in solving various EDA tasks. 
More importantly, GNNs are now employed to address several hardware security problems, such as detecting intellectual property (IP) piracy and hardware Trojans (HTs), to name a few. 

In this survey, we first provide a comprehensive overview of the usage of GNNs in hardware security and propose the first taxonomy to divide the state-of-the-art GNN-based hardware security systems into four categories: (i)~HT detection systems, (ii)~IP piracy detection systems, (iii)~reverse engineering platforms, and (iv)~attacks on logic locking. 
We summarize the different architectures, graph types, node features, benchmark data sets, and model evaluation of the employed GNNs. 
Finally, we elaborate on the lessons learned and discuss future directions.
\end{abstract}


\maketitle

\renewcommand{\headrulewidth}{0.0pt}
\thispagestyle{fancy}
\lhead{}
\rhead{}
\chead{\copyright~2022 IEEE.
This is the author's version of the work.
The definitive Version of Record is published in
2022 International Conference On
Computer-Aided Design (ICCAD)}
\cfoot{}

\section{Introduction}
\label{sec:Introduction}

\black{Integrated circuits (ICs) are ubiquitous in our daily lives. 
Not a day goes by where human beings do not interact with devices that do not have embedded ICs ranging from personal smartphones and laptops to automobiles. 
In order to extract every ounce of performance from these ICs, researchers continually employ the shrinking of the transistors and apply novel methods for developing transistors. 
However, the shrinking of technology nodes necessitates exorbitant investment~\cite{tsmc3nm} in building and commissioning state-of-the-art foundries, which has dissuaded design companies from operating their own foundries.\footnote{\black{Noticeable exceptions are Intel and Samsung, which have access to their foundries.}}
As a result, design companies typically outsource the fabrication of their designs to off-shore, third-party foundries, which could be potentially \textit{untrustworthy}.
}

\black{Outsourcing fabrication of ICs and other critical aspects of the supply chain (e.g., testing and packaging of ICs) enables untrustworthy entities to mount several hardware-focused threats.
These threats include piracy of the design intellectual property (IP), unauthorized overproduction of ICs, reverse engineering (RE), and implantation of malicious logic known as hardware Trojans (HTs)~\cite{rostami2014primer}.
Security researchers have proposed several defense strategies to combat the aforementioned threats, while
multiple concerted efforts have been made by researchers who find vulnerabilities 
by formulating attack strategies.
This game of cat-and-mouse between researchers from both spectrums (attackers and defenders) has led to the continual evolution of attacks and defenses, fuelling research in the hardware security community.
}

\black{On the other hand, graph neural networks (GNNs) have attracted considerable attention owing to their superior performance in graph-based learning applications~\cite{kipf2016semi,hamilton2017inductive,velivckovic2017graph}.
These applications include (but are not limited to) computer vision~\cite{xu2017scene}, traffic prediction~\cite{yao2018deep}, and recommender systems~\cite{ying2018graph}. Researchers have successfully utilized GNNs for several electronic design automation (EDA) tasks, such as floorplanning optimization~\cite{mirhoseini2021graph} and estimating routing congestion~\cite{kirby2019congestionnet}, to name a few. 
The success of GNNs in EDA is primarily because Boolean circuits can be naturally represented as graphs.
}
Recently, security researchers have incorporated GNNs into several
hardware security-related tasks and have demonstrated state-of-the-art performance in the detection of HTs~\cite{gnn4tj,yu2021hw2vec,hasegawa2021node,muralidhar2021contrastive,gnn4gate}, detection of IP piracy~\cite{yasaei2021gnn4ip,yu2021hw2vec}, reverse engineering of gate-level netlists~\cite{chowdhury2021reignn,gnnre}, unlocking hardware obfuscation~\cite{azar2020nngsat,gnnunlockp,omla,untangle}, and prediction of attack run-time on
logic locking~\cite{chen2020estimating}.

To the best of our knowledge, there is no published review on the rapidly developing topic of GNNs for hardware security. 
Our survey provides a comprehensive overview of GNN-based hardware security systems. 
The contributions of our work are as follows.

\begin{enumerate}[leftmargin=*]

\item \textbf{Comprehensive Review.} We present the first-of-its-kind survey on the role played by GNNs in investigating hardware security problems. 
To that end, we provide a comprehensive summary of the applicability of GNNs in downstream tasks ranging from reverse engineering, leaking secret keys from logic-locked designs, and detecting HTs.

\item \textbf{Abundant Resources.} We compile the resources on GNNs for hardware security, including models, input type, graph type, and features. 
This survey aids in understanding and employing different GNNs for various circuit-related tasks.

\item \textbf{Future Directions.} We outline the strengths and limitations of using GNNs for hardware security problems and discuss potential future directions.

\end{enumerate}

\noindent\textbf{Paper Organization.} The remainder of the paper is organized as follows. 
We provide an overview of selected hardware security threats arising from the globalization of the IC supply chain and selected design-for-trust solutions in Section~\ref{sec:background}.
Next, we provide suitable background regarding GNNs, including definitions for node and graph classification and link prediction, along with descriptions regarding common GNN architectures in Section~\ref{sec:GNNs}.
In Section~\ref{sec:GNN_for_hw_security}, we present our proposed taxonomy and provide details regarding the efforts undertaken by the research community in utilizing GNNs for different hardware security problems. Section~\ref{sec:discussion} provides a detailed discussion regarding the superiority and shortcomings of using GNNs for hardware security problems.
Finally, Sec.~\ref{sec:conclusion} presents concluding remarks.
\section{Background on Hardware Security}
\label{sec:background}

In this section, we provide background on some of the major threats in the domain of hardware security and provide details regarding selected countermeasures, \textit{i.e.,} \textit{design-for-trust} solutions.

\subsection{Hardware Security Threats}

\black{\textbf{Hardware Trojans (HTs)} are malicious modifications to ICs that attackers implant to achieve a malignant outcome~\cite{rostami2014primer}.
An HT comprises a trigger and payload. 
Attackers construct the trigger using nets in the design whose probability of activation is below some rareness threshold.\footnote{Rareness threshold is the probability below which nets are classified as rare nets.} 
The triggering condition determines the activation of HTs, and attackers design triggers as a combination of multiple rare events. 
The HT-implanted designs exhibit the intended functionality as HT-free designs until the triggering condition is satisfied.
HTs can leak information such as cryptographic keys, corrupt the intended functionality of an IC, and/or degrade the reliability of the underlying system.
In order to insert an HT in an IC, attackers are typically unrestricted, \textit{i.e.,} they can effect malicious modifications anywhere in the globalized IC supply chain.
For example, attackers in a third-party IP provider can augment additional logic gates before selling the IP to the design house~\cite{tehranipoor2010survey}.
On the other hand, attackers in an untrustworthy foundry can insert HTs by modifying the layout and mask information~\cite{tehranipoor2010survey}.
}

\noindent\black{\textbf{Reverse engineering}\footnote{\black{The legitimacy of reverse engineering has been established through the Semiconductor Chip Protection Act, which was adopted by the U.S. Government in 1984 and to which most industrialized countries adhere.}} of ICs can be categorized as (i)~product teardowns, (ii)~system-level analysis, (iii)~process analysis, and (iv)~circuit extraction~\cite{torrance2009state}.
Product teardowns help identify the internal board and components in an IC, while system-level analysis aids in analyzing the underlying functionality, interconnections between logic gates, etc.
Process analysis examines the structure and materials which could shed information about manufacturability aspects of an IC, whereas circuit extraction creates the underlying circuit-level schematics.
The reverse engineering of ICs serves multiple goals ranging from understanding the intrinsic details regarding a competitor's IC to detecting patent infringement by locating the stolen IP in the competitor's IC. 
Additionally, it can also help in detecting HTs and counterfeit products.
Several specialized companies, such as TechInsights~\cite{reverse_engineering}, perform on-demand reverse engineering as well.
}

\noindent\black{\textbf{IP piracy} is another threat which results due to the globalized IC supply chain. IP piracy refers to the theft of the design IP by an adversary (e.g., foundry, test facility, or end-user) to develop competing devices without incurring the costs involved in the research and development. 
The theft of design IP by a rogue nation can also lead to national security implications.
Hence, effective countermeasures are required to protect a design IP against piracy. In addition, detection techniques are also imperative to detect IP theft/piracy.
}

\subsection{Design-for-Trust Methods}

\textbf{Layout camouflaging} mitigates reverse-engineering attacks, \textit{i.e.,} reverse engineering of the underlying chip design IP performed by untrustworthy end-users~\cite{knechtel2019protect}. 
Camouflaging seeks to alter the appearance of a chip to conceal the design IP, \textit{i.e.,} it obfuscates the design information either at the transistor-level~\cite{patnaik2018advancing}, gate-level~\cite{rajendran2013security}, or interconnect-level~\cite{patnaik2020obfuscating}.
Based on the abstraction where camouflaging is performed, it can either be referred to as transistor camouflaging, gate camouflaging, or interconnect camouflaging.

\noindent\textbf{Split manufacturing} protects the design IP from untrustworthy foundries during fabrication~\cite{rajendran2013split,patnaik2021concerted}.
The design IP is split into front-end-of-line (FEOL) and back-end-of-line (BEOL).
Considered independently, the split portions become a ``sea of logic gates'' without any wiring (interconnect) information for the FEOL foundry, whereas it
becomes system-level wiring without any logic gate-related information for the BEOL foundry.
Split manufacturing is beneficial since: (i)~outsourcing of the FEOL requires access to advanced technology nodes, and (ii)~BEOL fabrication is significantly less complex than FEOL fabrication.

\begin{figure}[!t]
\centering
\includegraphics[width=0.38\textwidth]{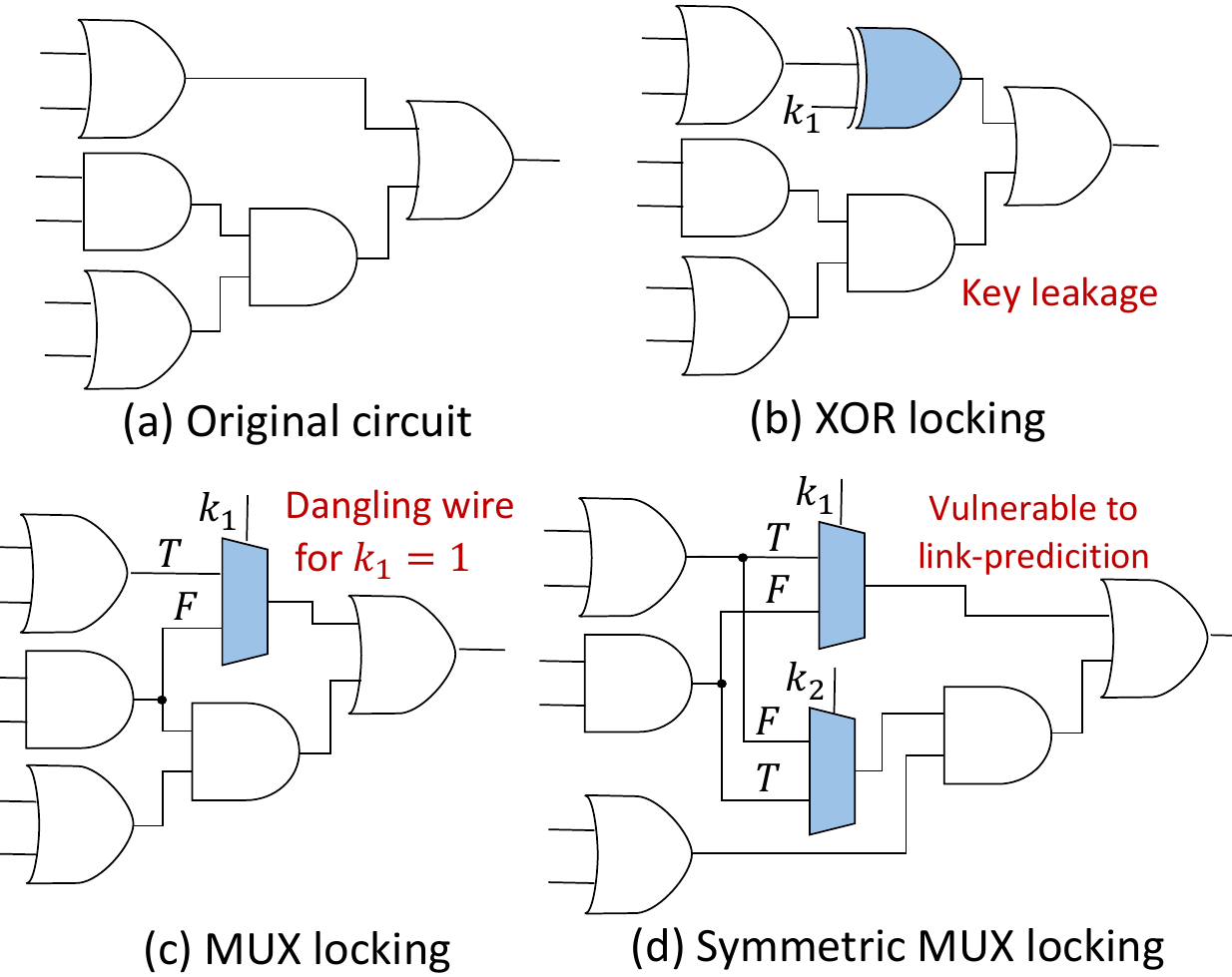}
\smallerspacecaption
\caption{Examples of logic locking. T and F denote true and false connection, respectively.}
\label{fig:example}
\end{figure}

\noindent\textbf{Logic locking} obfuscates the structure and functionality of a design by integrating key-controlled logic elements,
referred to as \textit{key-gates}. These key-gates bind the correct functionality of the design to a secret key that
is only known to the legitimate IP owner. Thus, during outsourced design and fabrication stages, the design is not revealed in
full. In coordination with other trusted parties, the owner loads the secret key into an on-chip, tamper-proof memory
after fabrication and testing. Figure~\ref{fig:example}~(a) demonstrates an example circuit, which is locked
using an XOR key-gate in~(b) versus using a multiplexer (MUX) in~(c). The secret key in both cases is $k_1=0$.

Researchers evaluate the security of logic locking mainly under two threat models, \textit{i.e.,}
	 \textit{oracle-guided}~\cite{Subramanyan_host_2015} and \textit{oracle-less} attack models~\cite{sirone2020functional}. In the oracle-guided model, an adversary has access to a functional chip holding the key, an \textit{oracle}. In contrast, the oracle-less model relies only on the netlist structure of the locked design to decipher the secret key~\cite{redundancy} or to
remove the protection logic altogether~\cite{removal,limaye2022valkyrie}. \textit{In the past few years, researchers started to focus on the oracle-less model}, which is more
realistic and powerful.

\section{Background on Graph Neural Networks (GNNs)}
\label{sec:GNNs}

In this section, we provide a background on GNNs, define important graph-related concepts, and depict the notations used in this paper (Table~\ref{tab:symbol}).
We begin by defining a graph as follows.
\begin{definition}
\textbf{(Graph)}, $\gG=(V,E)$ denotes a graph with set $V$ of nodes and set $E \subseteq V\times V$ of edges. $\mX \in \R^{n \times k}$ is a matrix of node features, where $n$ is the number of nodes in $\gG$. $\mA \in \{0,1\}^{n\times n}$ represents the adjacency matrix of $\gG$ with $\mA_{i,j}=1$ iff $(i,j)\in E$. 

\end{definition}

\begin{definition}
A \textbf{(directed graph)} is a graph in which the edges have a direction from one node to another.
\end{definition}
\begin{definition}
An \textbf{(undirected graph)} is a special case of directed graphs where there is a pair of edges with opposite directions for every two connected nodes. $\mA$ of an undirected graph is symmetric.

\end{definition}

\begin{figure}[tb]
\includegraphics[width=0.7\columnwidth]{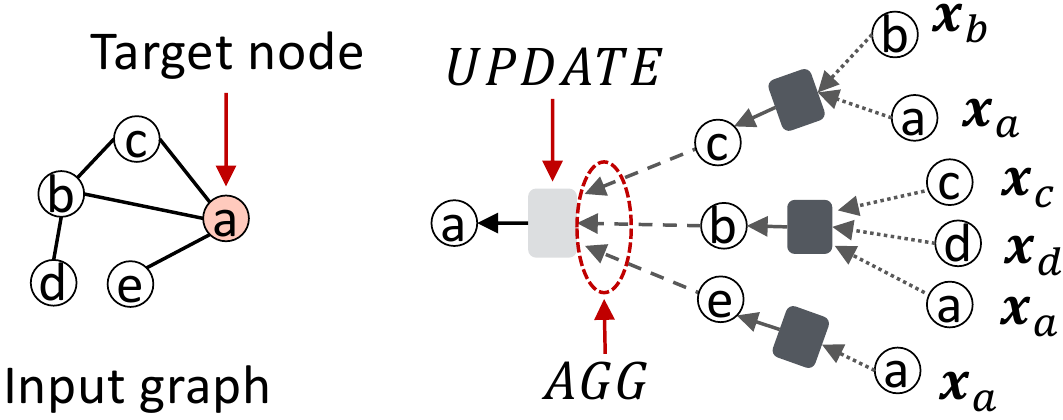}
\caption{In GNNs, the embedding of a node is updated after aggregating the features of its neighbors~\cite{gnnre}.}
\label{fig:gnn_aggregate}
\end{figure}

A GNN generates a vector representation (\textit{embedding}) for each node in the graph such that similar nodes are placed together in the embedding space.
The embedding of a target node $v$ gets updated through \textit{message passing} (\textit{neighborhood aggregation}), as illustrated in Figure~\ref{fig:gnn_aggregate}. 
The features of the neighboring nodes are accumulated to generate an aggregated representation. 
The aggregated information is then combined with the features of the target node to update its embedding. 
Consequently, after $L$ rounds of message passing, each node is aware of its features, the features of the neighboring nodes, and the structure of the graph within the $L$-hop neighborhood. 
The message passing phase is abstracted as follows, where $\mz_v^{(l)}$ indicates the embedding of node $v$ at the $l$-th round, $\mz_v^{(0)}=\mx_v$, and $\gN(v)$ represents the first-order neighbors of node $v$.
\begin{align}
\vspace*{-6mm}
 \ma_v^{(l)} = \mathsf{AGG}^{(l)} \left( \left\lbrace \mz_u^{(l-1)} : u \in \gN(v) \right\rbrace \right) 
\vspace*{-6mm}
\end{align}
\vspace*{-5mm}
\begin{align}
\vspace*{-5mm}
 \mz_v^{(l)} = \mathsf{UPDATE}^{(l)} \left( \mz_v^{(l-1)}, \ma_v^{(l)} \right)
 \vspace*{-5mm}
\end{align}

GNNs mainly differ based on the choices of the $\mathsf{AGG}(\cdot)$ and $\mathsf{UPDATE}(\cdot)$ functions. The $\mathsf{AGG}(\cdot)$ function is typically an order invariant function, such as $\mathsf{sum}$, $\mathsf{average}$, or $\mathsf{max}$. 

\subsection{Node and Graph Classification}
\begin{definition}

\textbf{(Node classification)}, each node $v \in V$ is associated with a true label $y_v$ and the GNN learns an embedding $\mz_v^{(L)}$ of $v$ such that $v$'s label can be predicted as $\hat{y}_v=g(\mz_v^{(L)})$. Where $g$ is a downstream classifier.
\end{definition}
\begin{definition}
\textbf{(Graph classification)}, given a set of graphs\\
$\{\gG_1, ..., \gG_N\}$ and their labels $\{y_1, ..., y_N \}$, the GNN learns an embedding $\mz_\gG$ such that $\gG$'s label can be predicted as, $\hat{y}_\gG=g(\mz_\gG)$.
\end{definition}

The generated node embeddings can be directly used for node classification. For graph classification tasks, a $\mathsf{readout}$ function is performed to generate a graph-level embedding, $\ssub{z}{\gG}$, which can be used for graph classification.

\begin{table}[tb]{\footnotesize
\renewcommand\arraystretch{0.9}
\centering
\caption{Symbols and notations used in this work \label{tab:symbol}}}
\resizebox{0.5\textwidth}{!}{%
\begin{tabular}{llll}
\hline
\textbf{Notation} & \textbf{Definition} & \textbf{Notation} & \textbf{Definition} \\
\hline
$\gG, \ssub{y}{G}$ & Graph, class & $\mX$ & Node features matrix \\\hline
$v, \ssub{y}{v}$ & Node, class & $\mz_{\gG}$ & Graph embedding \\\hline
 $\mA$ & Adjacency matrix & $h$ & hop-size \\\hline
 $V$ & Set of nodes in $\gG$ & $T$, $S$ & Sets of target links and nodes \\\hline
 $E$ & Set of edges in $\gG$ & $d(u,v)$ & Shortest distance between $u$ and $v$ \\\hline
 $n$ & Number of nodes in $\gG$ & $\gG_{(S,h)}$ & Subgraph from $\gG$ around $S$ \\\hline
 $k$ & Length of feature vector & $\hat{y}$ & Predicted label \\\hline
 $\gN(v)$ & Neighbors of nodes $v$ & $g$ & Downstream classifier \\\hline
 $L$ & Number of GNN layers & $\bm{W}_{(l)}$, $\bm{B}_{(l)}$& GNN trainable parameters \\\hline
 $\mZ$ & Node embeddings matrix & $\sigma(.)$ & Activation function \\\hline
\end{tabular}}
\end{table}

\begin{figure*}[tb]
\centering
\includegraphics[width=0.85\textwidth]{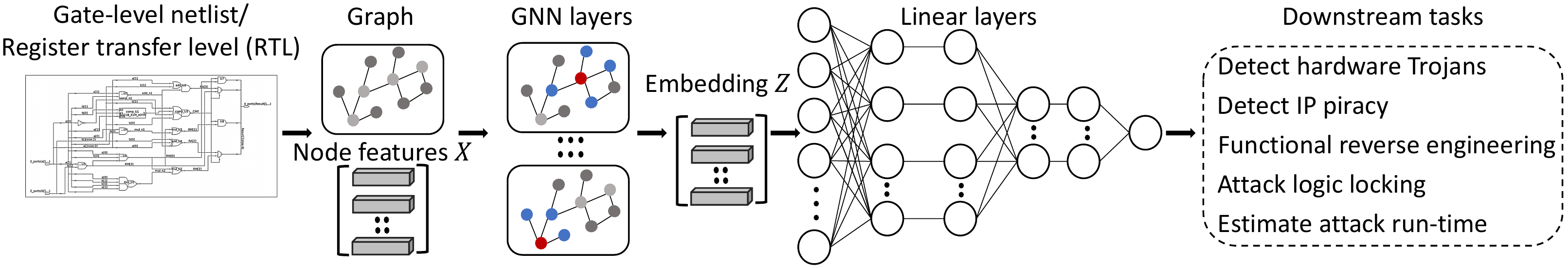}
\caption{Applications of GNNs in hardware security.}
\label{fig:gnn_model}
\end{figure*}

\subsection{Link Prediction Problem}

Link prediction is to estimate the likelihood of the link between two nodes in a given network based on the structure of the network and the properties of its nodes~\cite{liben2007link}.
Link prediction can be used for recommendation systems~\cite{adamic2003friends}, protein interaction prediction~\cite{qi2006evaluation}, drug response prediction~\cite{stanfield2017drug}, and many other applications.

Link prediction algorithms assign \textit{likelihood scores} $\in [0, 1]$ to links of interest (\textit{target links $T$}), where $T \notin E$. A target link is denoted by the node set $S\in V$, which includes the two end nodes of the link. Traditional link-prediction scoring is based on some predefined heuristics, such as the number of common neighbors or the degree of the nodes. However, the selection of such heuristics requires manual efforts, depending on the type of graph.

Recently, GNNs have shown tremendous success in performing link prediction, exploiting both the structure of the graph and the associated node features to extract link features, surpassing the performance of traditional methods~\cite{SEAL}. In GNN-based link prediction, an \textit{enclosing subgraph} around each target link is extracted.

\begin{definition}

\textbf{(Enclosing subgraph)}, given $(S,\gG)$, the $\gG_{(S,h)}$ subgraph induced from $\gG$ by $\cup_{v\in S} \{u ~|~ d(u,v) \leq h\}$, where $d(u,v)$ is the shortest path distance between nodes $u$ and $v$.

\end{definition}

The enclosing subgraphs hold information about the network surrounding the links. Therefore, by performing graph classification, the labels of the target links also become the labels of their corresponding subgraphs.

\subsection{Common GNN Architectures}
Next we provide details for some of the commonly used GNNs.

\subsubsection{GraphSAGE}
The fundamental mean aggregator function in GraphSAGE~\cite{hamilton2017inductive} is described below.

\begin{gather}
\mz_v^{(l)}=\sigma ([\bm{W}_{(l)}\cdot \mathsf{AGG}(\{\mz_u^{(l-1)},\forall u\in \gN(v)\}),\bm{B}_{(l)} \mz_v^{(l-1)}])\\
\mathsf{AGG}=\sum_{u\in \gN(v)}\frac{\mz_u^{(l-1)}}{|\gN(v)|}
\end{gather}

where $\sigma(.)$ is an activation function, $\bm{W}_{(l)}$ and $\bm{B}_{(l)}$ are trainable matrices (\textit{i.e.,} encoding what the model learns). GraphSAGE concatenates the self-embedding of the previous step $\mz_{v}^{(l-1)}$ with the neighbor embedding (output of the $\mathsf{AGG}$
function) in order to update the embedding of a node $\mz_{v}^{(l)}$. The $\bm{W}_{(l)}$ transformation learns the important
components from the neighbors' features and the $\bm{B}_{(l)}$ transformation learns the important features of the node itself. Note that
individual weight parameters are defined for each transformation and per layer $l$.

\subsubsection{Graph Convolutional Network (GCN)} A single GCN layer~\cite{kipf2016semi} is as follows:
\begin{align}
 \vspace*{-5mm}
 \mZ^{(l+1)}=\sigma(\tilde{\mD}^{-1}\tilde{\mA}\mZ^{(l)}\mW_{(l)})
 \vspace*{-5mm}
\end{align}

where $\tilde{\mA}=\mA +\mI$ adds self loops to allow self aggregation. $\tilde{\mD}$ is the diagonal degree matrix, where $\tilde{\mD}_{ii}=\sum_j \tilde{\mA_{i,j}}$, and $\mW_{(l)}\in\R^{k_{(l)} \times k_{(l+1)}}$ is a trainable weight matrix. $\sigma(.)$ is an element-wise non-linear activation function. 
$\mZ^{(l+1)}\in\R^{n \times k_{(l+1)}}$ is the output embedding of layer $l$, where $\mZ^{(0)}=\mX$.
The first step in the convolutional layer is $\mZ^{(l)}\mW_{(l)}$, which performs a linear feature transformation on node information, mapping the $k_{(l)}$ feature channels to $k_{(l+1)}$ channels. 
The second step aggregates the node information to neighboring nodes, including the node itself. 
Then $\tilde{\mD}$ normalizes the aggregated information to ensure a fixed feature scale.
Multiple convolutional layers can be employed to extract multi-scale sub-structure features from the network.

\subsubsection{Deep Graph Convolutional Neural Network (DGCNN)} The DGCNN architecture~\cite{zhang2018end} achieves superior results in graph classification. After $L$ GCN layers, the output embeddings from each layer $l=1,\ldots,L$ are concatenated horizontally, to capture the graph in a single output vector $\mZ^{(1:L)}:=[\mZ^{(1)}, \ldots, \mZ^{(L)}]$, where $\mZ^{(1:L)}\in\R^{n \times \sum_{l=1}^{L} k_{(l)}}$. 
A \textit{sort pool layer} takes in the $n\times\sum_{l=1}^{L}k_{(l)}$ tensor $\mZ^{1:L}$ and sorts it row-wise according to $\mZ^{L}$. 
The final tensor is reshaped to $c(\sum_{l=1}^{L} k_{(l)}) \times 1$, selecting $c$ nodes to represent the graph. 
Then, the final embedding is fed to $1$-D convolutional layers with filter and step size of $\sum_{l=1}^{L}k_{(l)}$ to classify the graph.

\subsubsection{Graph Isomorphism Network (GIN)} The GIN architecture~\cite{xu2018powerful} is one of the most expressive GNNs. 
GIN updates the node embeddings as follows, where MLP represents a multi layer perceptron.
\begin{align} \label{GIN-agg}
\centering
\mz_v^{(l)} = {\rm MLP}^{(l)} \left( \mz_v^{(l-1)} + \sum\nolimits_{u \in \mathcal{N}(v)} \mz_u^{(l-1)}\right)
\end{align}
GIN considers all structural information from all iterations of the model by replacing the traditional $\mathsf{readout}$ functions with subgraph embeddings concatenated across \emph{all layers} of GIN as follows. The $\mathsf{readout}$ function adds all node embeddings from the same layer. 
\begin{align}
\label{eq:GIN-readout}
\centering
\mz_{\gG} = \func[CONCAT]{ \func[READOUT]{ \left\lbrace \mz_v^{(l)} | v \in \gG \right\rbrace} \ \big\vert \ l = 0,1, \ldots, L }
\end{align}

\subsubsection{The Graph Attention Network (GAT)} GAT~\cite{velivckovic2017graph} measures the importance of the edges during aggregation. It employs a multi-head attention method of $K$, in which the layer $l$'s information propagates to layer $l+1$ as follows;
\begin{equation}
{\mz}_{v}^{(l+1)}=
\left|\begin{matrix}
\\ 
\\ 
\end{matrix}\right|_{k=1}^K \sigma\begin{pmatrix}
\sum_{u\in \gN(v)} \alpha_{u,v}^{(k)} {W}^{(k)} {\mz}_{v}^{l}

\end{pmatrix}
\\
\end{equation}
\begin{equation}
 \alpha_{u,v}^{(k)}={LeakyReLU}
\begin{pmatrix}
({\ma}^{(k)})^\intercal[{W}^{(k)}{h}_u \parallel {W}^{(k)} {\mz}_v]
\end{pmatrix}
\end{equation}
Where $\alpha_{u,v}$ specifies the weighting factor of node $u$'s features for node $v$, which is computed as a byproduct of an attention mechanism ${\ma}$. The multi-head attention mechanism replicates the aggregation layers $K$ times, each replica having different trainable parameters ${W}^{(k)}$, and the outputs are feature-wise aggregated using a concatenation operation as described in Equation~(8).

\section{GNNs for Hardware Security}
\label{sec:GNN_for_hw_security}

In this section, we outline the applications of GNNs in the field of hardware security and summarize them in Figure~\ref{fig:gnn_model}. 
In Table~\ref{tab:p_taxonomy}, we present our proposed taxonomy, grouping the GNN-based hardware security systems into four categories: (i)~hardware Trojan detection systems, (ii)~IP piracy detection systems, (iii)~reverse engineering platforms, and (iv)~attacks on logic locking. 
We discuss each category in the following sub-sections. 
Further, in Table~\ref{tab:my-table}, we summarize the main aspects of representative GNN for hardware security platforms concerning: (i)~GNN tasks, (ii)~GNN architectures, (iii)~number of GNN layers, (iv)~loss functions, (v)~graph type, (vi)~circuit type, (vii)~features, (viii)~pooling layers, and (ix)~readout layers.

\begin{table}[tb]
\caption{Proposed taxonomy and representative publications of GNNs for hardware security}
\label{tab:p_taxonomy}
\resizebox{0.7\columnwidth}{!}{%
\begin{tabular}{ll}
\hline
\textbf{Category} & \textbf{Publications} \\ \hline
\textbf{Hardware Trojan detection} & ~\cite{gnn4tj,gnnre,yu2021hw2vec,hasegawa2021node,muralidhar2021contrastive,gnn4gate} \\ \hline
\textbf{IP piracy detection} & ~\cite{yu2021hw2vec,gnn4ip,gnnre} \\ \hline
\textbf{Reverse engineering} & ~\cite{gnnre,chowdhury2021reignn,wang2022functionality,he2021graph,zhao2022graph} \\ \hline
\textbf{Attacks on logic locking} & ~\cite{omla,muxlink,untangle,gnnunlockp,alrahis2020gnnunlock,chen2020estimating} \\ \hline
\end{tabular}%
}
\end{table}

\begin{table*}[tb]
\caption{Summary of GNNs for hardware security. ``$-$'' means unspecified}
\label{tab:my-table}
\resizebox{\textwidth}{!}{%
\setlength\tabcolsep{1.9pt} 
\renewcommand\arraystretch{0.9}
\begin{tabular}{llllllllll}
\hline
\textbf{Platform} & \textbf{GNN Task} & \textbf{GNN} & \textbf{\#Layers} & \textbf{Loss Function} & \textbf{Graph Type} & \textbf{Circuit Design Level} & \textbf{Features} & \textbf{Pooling} & \textbf{Readout} \\ \hline
\textbf{GNN4TJ~\cite{gnn4tj}} & Graph classification & GCN~\cite{kipf2016semi} & 2 & Cross-entropy & DFG & RTL/gate-level netlist & Type of node operation & Top-k filtering. & Max \\ \hline
\textbf{GNN4IP~\cite{gnn4ip}} & Graph similarity & GCN~\cite{kipf2016semi} & 2 & Cosine-similarity & DFG & RTL/gate-level netlist & Type of node operation & Top-k filtering & Max \\ \hline
\textbf{GNN-RE~\cite{gnnre}} & Node classification & GraphSAINT~\cite{zeng2019graphsaint} & 5 & Cross-entropy & Undirected & Gate-level netlist & \begin{tabular}[c]{@{}l@{}}Input degree\\ Output degree\\ Connectivity to ports\\ Gate type\\Neighborhood info\end{tabular} & Multi-head attention & NA \\ \hline

\textbf{ABGNN~\cite{he2021graph}} & Node classification & ABGNN~\cite{he2021graph}& 1-5-2 & Cross-entropy & Directed & Gate-level netlist & $-$ & NA & NA \\\hline

\begin{tabular}[c]{@{}l@{}}
\textbf{GNN-RE under}\\
\textbf{circuit rewriting~\cite{zhao2022graph}}
\end{tabular}
 & Node classification & GCN~\cite{kipf2016semi} & 3 & $-$ & DAG & Gate-level netlist & \begin{tabular}[c]{@{}l@{}}Input degree\\ Output degree\\ Connectivity to ports\\ Gate type\\Neighborhood info\\Truth table\\Signal probability\\NPN class\end{tabular}& NA & NA \\\hline

\textbf{NCL~\cite{wang2022functionality}} & \begin{tabular}[c]{@{}l@{}}Node classification\\Graph classification\end{tabular} & FGNN~\cite{wang2022functionality} & $-$ & Cross-entropy & Directed & Gate-level netlist & $-$ & NA & Mean \\\hline

\textbf{ReIGNN~\cite{chowdhury2021reignn}} & Node classification & GraphSAGE~\cite{hamilton2017inductive} & 3 & 
\begin{tabular}[c]{@{}l@{}}Negative \\ log-likelihood\end{tabular}

 & $-$ & Gate-level netlist & \begin{tabular}[c]{@{}l@{}}Input degree\\ Output degree\\ Harmonic centrality\\ Gate type \\ Betweenness centrality\\ Neighborhood info\end{tabular} & NA & NA \\\hline

\textbf{ICNet~\cite{chen2020estimating}} & Graph regression & GAT~\cite{velivckovic2017graph} & 2 & \begin{tabular}[c]{@{}l@{}}Mean squared\\ error (MSE)\end{tabular} & Undirected&Gate-level netlist & \begin{tabular}[c]{@{}l@{}} Gate type\\
Key-gate mask\end{tabular} & Attention & Attention \\\hline

\begin{tabular}[c]{@{}l@{}}
\textbf{GNNUnlock~\cite{alrahis2020gnnunlock}}\\
 \textbf{GNNUnlock+~\cite{gnnunlockp}}
\end{tabular}
 & Node classification & GraphSAINT~\cite{zeng2019graphsaint} & 2 & Cross-entropy & Undirected & Gate-level netlist & \begin{tabular}[c]{@{}l@{}}Input degree\\ Output degree\\ Connectivity to ports\\ Gate type\end{tabular} & NA & NA \\ \hline
\textbf{OMLA~\cite{omla}} & Subgraph classification & GIN~\cite{xu2018powerful} & 5-6 & Cross-entropy & Undirected & Gate-level netlist & \begin{tabular}[c]{@{}l@{}}Gate type\\ Connectivity to ports\\ Distance encoding\end{tabular} & \begin{tabular}[c]{@{}l@{}}Summing node features\\ from the same layer\end{tabular} & Sum \\ \hline

\begin{tabular}[c]{@{}l@{}}
\textbf{MuxLink~\cite{muxlink}}\\
\textbf{UNTANGLE~\cite{untangle}}

\end{tabular}

 & Link prediction & DGCNN~\cite{zhang2018end} & 4 & Cross-entropy & Undirected & Gate-level netlist & \begin{tabular}[c]{@{}l@{}}Gate type\\ Distance encoding\end{tabular} & DGCNN sortpooling~\cite{zhang2018end} & Max \\ \hline
\end{tabular}%
}
\end{table*}

\subsection{GNNs for Hardware Trojan Detection}
\label{sec:GNNs_HT_detection}

Third-party IPs (3PIPs) in 	register transfer level (RTL) format are complex and flexible, supporting multiple configurations for different applications, which is a convenient structure for adversaries to insert HTs.
In the case of untrusted 3PIPs, a golden model (\textit{i.e.,} HT-free) of the IP is unavailable, and thus, it is challenging to detect possible HTs using testing-based~\cite{hicks2010overcoming} or side-channel-based methods~\cite{huang2018scalable}. 
Destructive methods (\textit{i.e.,} depackaging, delayering, and reverse engineering, followed by a circuit-level comparison~\cite{kommerling1999design,nohl2008reverse}) can check if ICs are HT-infected but only after fabrication when the damage is already done~\cite{bao2015reverse}. 
Other HT detection methods (e.g., graph-similarity-based techniques~\cite{fyrbiak2019graph}) have several shortcomings, such as complexity~\cite{chakraborty2008demand} and the inability to identify unknown HTs.

GNN4TJ~\cite{gnn4tj} is a GNN-based platform that detects HTs without requiring prior knowledge of the design IP or HT structure. 
GNN4TJ converts the RTL design into a corresponding data flow graph (DFG). This DFG is then fed to a GNN to extract features and learn the structure and behavior of the underlying design. 
Subsequently, the GNN performs a graph classification task and assigns a label $\hat{y}$ to each design $p$ based on the presence of HTs. The GNN learns the properties of HTs and generalizes them to unseen HTs. Researchers have proposed other GNN-based platforms for HT detection~\cite{hasegawa2021node,muralidhar2021contrastive,gnn4gate}, extending the task to node classification to locate the HT.

\subsection{GNNs for IP Piracy Detection}
\label{sec:background_IP}

GNN4IP~\cite{gnn4ip} is a GNN-based IP piracy detection technique that assesses the similarity between circuits revealing potential theft.
In GNN4IP, the structure of the design IP becomes its signature. 
Hence, GNN4IP does not require the addition of any watermarks or fingerprints (thereby reducing overheads) that could be prone to removal attacks~\cite{alkabani2007remote,cui2015ultra}.
GNN4IP compares two circuits ($p_1$ and $p_2$) either in RTL or gate-level logic representation. 
Like GNN4TJ, the circuits are converted to DFG or abstract syntax tree (AST) format and fed to a GNN. 
The GNN generates an embedding for each circuit from its underlying structure (\textit{i.e.,} signature). 
Subsequently, the GNN optimizes the embeddings so that distances in the embedding space reflect the similarity between designs (\textit{i.e.,} graphs)~\cite{hamilton2017inductive}.
Therefore, GNN4IP infers piracy by computing the \textit{cosine similarity score} between the obtained embeddings as follows, where $z_{p_1}$ and $z_{p_2}$ represent the embedding vectors of designs $p_1$ and $p_2$. 
Finally, GNN4IP compares the similarity score with a predefined decision boundary $\delta$ to predict whether there is piracy between the two circuits, returning a binary label as its output ($0$ or $1$).

\begin{equation}
\vspace{-0.2em}
   \text{cosine\_sim}(z_{p_1}, z_{p_2}) =
    \frac{z_{p_1} \cdot z_{p_2}}{|z_{p_1}||z_{p_2}|}
\end{equation}

HW2VEC~\cite{yu2021hw2vec} combines the GNN4TJ~\cite{gnn4tj} and the GNN4IP~\cite{gnn4ip} platforms into a single framework, handling RTL designs and gate-level netlists. In addition, HW2VEC supports different graph types, such as DFG and abstract syntax tree (AST).

\subsection{Functional Reverse Engineering}

Prior art in functional RE typically applies the following work-flow. 
A set of candidate sub-circuits is extracted (e.g., by partitioning the netlist), and then each sub-circuit is labeled (e.g., by performing exhaustive formal verification against components from a golden library)~\cite{li2013wordrev,li2012reverse,subramanyan2013reverse,gascon2014template}.
These works have the following limitations: (i)~extracting all relevant candidate sub-circuits and checking each candidate by formal verification is time-consuming,\footnote{The performance and accuracy of such an approach depend heavily on the constructed golden library.} and (ii)~such techniques cannot identify any variants of design components in the golden library~\cite{baehr2020machine}. 

Recently, GNN-based approaches have been proposed to advance
functional RE of digital circuits, as discussed below.

In the GNN-based functional RE platform, GNN-RE, the gate-level netlist of a circuit is first transformed into an undirected graph representation, preserving the structure of the netlist, \textit{i.e.,} gate connectivity~\cite{gnnre}. 
Each gate in the netlist is represented as a node in the corresponding graph and initialized with a feature vector that encodes its gate type (\textit{i.e.,} XOR, XNOR, AND, OR, etc.), its neighboring gates, input degree, output degree, and connectivity to ports (\textit{i.e.,} primary inputs or primary outputs). GNN-RE then performs a node classification task, \textit{i.e.,} identifying which gates belong to which sub-circuit. 
On datasets of circuits with several sub-circuits (multipliers, adders, finite state machines, etc.), GNN-RE achieves a node classification accuracy higher than $80\%$.

Similarly, in ReIGNN~\cite{chowdhury2021reignn}, the gate-level netlist is represented as a graph. A GNN processes the corresponding graph representation to discriminate between state and data registers in the netlist.

The asynchronous bidirectional GNN (ABGNN) platform analyzes gate-level netlists for sub-circuit classification, focusing on arithmetic blocks, specifically adders~\cite{he2021graph}. ABGNN represents the gate-level netlists as directed graphs to maintain the natural directed representation of circuits. ABGNN predicts the boundaries of adder blocks by training two separate GNNs, one for $\gG$ and one for $\gG^{\intercal}$. 
In other words, one GNN aggregates information from predecessors and the other from successors. The generated two embeddings are then combined as the final embedding.

All the aforementioned methods focus on the structural properties of the circuits for prediction. 
Authors in~\cite{wang2022functionality} proposed a contrastive learning (CL)-based netlist representation learning (NCL) framework to extract the logic functionality, which is universal and transferable across different circuits. 
The authors also propose the functionality graph neural network (FGNN) that works with the NCL platform for classification purposes, considering the Boolean functionality in addition to the structure.

FGNN is inspired by ABGNN~\cite{he2021graph}. However, it uses independent aggregators for nodes of different types.

Addressing the same problem, the authors in~\cite{zhao2022graph} employ a GNN for functional RE under circuit rewriting. 
The authors consider the features used in the GNN-RE platform in addition to a set of functionality-aware features, improving the detection accuracy. 
The authors considered features such as selected truth table information, signal probability, and negation-permutation-negation (NPN) class.

\subsection{Evaluating the Security of Logic Locking}

State-of-the-art GNN algorithms have demonstrated strong results in learning the structure of locked
circuits and breaking logic locking in an oracle-less
setting~\cite{ml_locking_survey,gnnunlockp,omla,untangle,muxlink}. 
The usage of GNNs exposed the following shortcomings.

\noindent\textbf{Key leakage.} In traditional X(N)OR locking~\cite{epic_journal,yasin_TCAD_2016,JV-Tcomp-2013}, there is a
direct mapping between the key-gates and the corresponding key-bit values: XOR key-gates require key-bit `0' whereas
XNOR key-gates require key-bit `1'. Re-synthesis runs can serve well for logical and structural transformations, with
the aim of obfuscating this correlation. Nevertheless, machine learning (ML) attacks such as SAIL~\cite{sail}, SnapShot~\cite{snapshot}, and OMLA~\cite{omla} learn such synthesis-induced modifications and to still decipher the key-value.

OMLA is a GNN-based attack on X(N)OR locking. OMLA first represents the locked gate-level netlist as an undirected graph. Then, OMLA extracts an h-hop enclosing subgraph around each key-gate. The subgraphs include information regarding the key-gates and their surrounding circuitry. Therefore, OMLA deciphers the key-bit values by performing a subgraph classification task using a GNN.

OMLA assigns each node in the extracted subgraph a feature vector that captures its Boolean function and connectivity to ports. In addition to the gate information, OMLA proposes a distance encoding method in which each node is assigned a label (tag) that indicates the shortest path distance from the node to the target key-gate in the subgraph. With such a labeling method, the target key-gate will always get a unique label of $0$, allowing the GNN to distinguish the key-gate from the rest of the nodes.

\noindent\textbf{Link formation.} Deceptive MUX (D-MUX)~\cite{sisejkovic2021deceptive} and symmetric MUX locking~\cite{alaql2021scope} circumvented all existing ML attacks prior to their development. Both schemes employ MUX
key-gates to eradicate key leakage and carefully select T and F wire to eliminate structural flaws, as depicted in
Figure~\ref{fig:example}~(d). 

However, the GNN-based MuxLink~\cite{muxlink} attack exposed a new vulnerability, \textit{i.e.,} link formation, breaking both D-MUX and symmetric MUX locking schemes. 
The intuition behind MuxLink is that modern ICs contain a large amount of repetition and reuse cores~\cite{saha2011soc}. 
MuxLink employs a GNN to learn the remaining (intact) structure of the locked design and then makes predictions regarding obfuscated interconnects. 
More specifically, MuxLink converts the problem of deciphering the inputs of a locking MUX to a link-prediction problem and solves it using a GNN.

\begin{figure}[tb]
\centering
\includegraphics[width=0.45\textwidth]{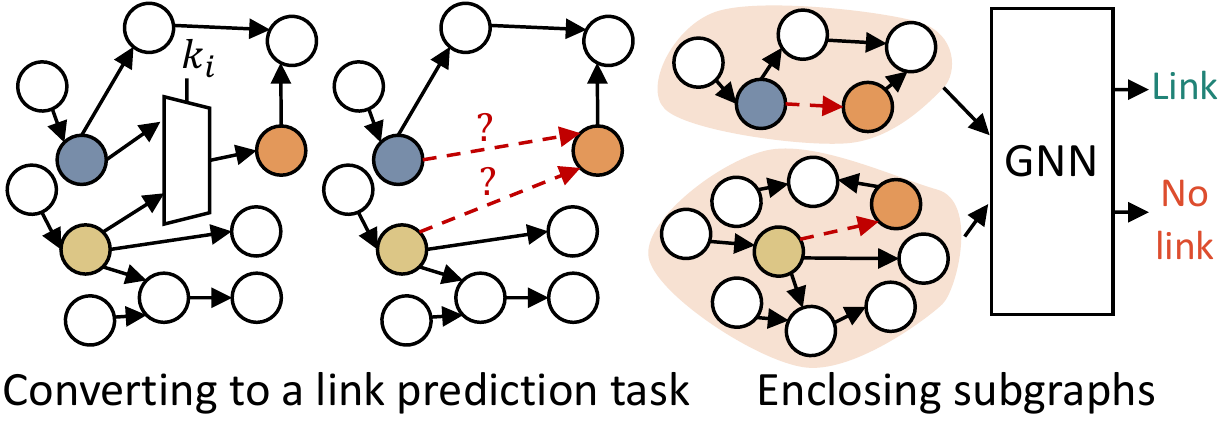}
\smallerspacecaption
\caption{MuxLink high-level concept~\cite{muxlink}.}
\label{fig:muxlinkdetails}
\end{figure}

Considering the example in Figure~\ref{fig:muxlinkdetails}, MuxLink converts the MUX locked locality into a graph with missing connections. 
In this example, two different connections are possible depending on the configuration of key-input. 
MuxLink then extracts h-hop enclosing subgraphs around the target links. MuxLink trains a GNN on the remaining unobfuscated interconnects in the design, learn the composition of gates, and then makes predictions on the target subgraphs. 
The GNN predictions are then processed, deciphering the secret key. 
MuxLink is successful because each of the extracted subgraphs is different and will have different likelihood values. 
Similarly, the GNN-based UNTANGLE~\cite{untangle} attack applies the same concept to unlock the state-of-the-art routing obfuscation scheme, InterLock~\cite{InterLock}. 
After GNN prediction, UNTANGLE employs a post-processing step to iteratively decipher the obfuscated connections due to the complexity of the targeted InterLock locking scheme. 
Such a post-processing scheme is not required for MuxLink when attacking D-MUX or random MUX locking.

\noindent\textbf{Structural leakage.} GNNUnlock~\cite{gnnunlockp,alrahis2020gnnunlock} is an attack on provably secure logic locking (PSLL) techniques that leverage a GNN to identify and isolate the protection logic. Generally, in PSLL techniques, the protection logic is embedded in the design. 
Identifying the protection logic facilitates its removal, enabling the recovery of the original design~\cite{removal,limaye2022valkyrie}.
PSLL techniques are different in terms of their construction compared to the traditional logic locking (TLL) solutions, considered by OMLA and MuxLink.

In TLL, the key-gates are spread across the design, and removing them will alter the functionality of the design. 
Hence, the corresponding key-bits must be deciphered to unlock the design functionality.
Unlike OMLA and MuxLink, GNNUnlock is a removal attack and is not applicable for TLL solutions.
OMLA/MuxLink and GNNUnlock are complementary attacks and can be used to circumvent compound locking techniques, where a PSLL technique is incorporated with a TLL (e.g., random logic locking) to achieve two-layered protection.
Note that, apart from the use-case applications, there are also differences concerning the usage of the GNN. 
OMLA and MuxLink perform a subgraph classification task, whereas GNNUnlock performs a node classification task. 
OMLA leverages the GIN architecture~\cite{xu2018powerful} and MuxLink leverages the DGCNN architecture~\cite{zhang2018end}, while GNNUnlock utilizes GraphSAINT~\cite{zeng2019graphsaint}.
OMLA and MuxLink employ a node labeling method, which is not required for GNNUnlock. 
In summary, the goal of the attacks, the GNN models, the classification tasks, the outcomes, and the inner workings of the attacks are distinct from each other.
Table~\ref{tab:compare_response} summarizes the differences between the GNN-based attacks on logic locking.

\begin{table}[tb]
\caption{Summary of GNN-based attacks on logic locking}
\centering
\label{tab:compare_response}
\resizebox{0.45\textwidth}{!}{%
\begin{tabular}{lllll}
\hline
\textbf{Attack} & \textbf{GNN Model} & \textbf{GNN Task} & \textbf{\begin{tabular}[c]{@{}c@{}}Attack\\ Type\end{tabular}} & \textbf{\begin{tabular}[c]{@{}c@{}}Attacked\\ Schemes\end{tabular}} \\ \hline
\textbf{GNNUnlock~\cite{gnnunlockp,alrahis2020gnnunlock}} & GraphSAINT~\cite{zeng2019graphsaint} & Node Classification & Removal & PSLL \\ \hline
\textbf{OMLA~\cite{omla}} & GIN~\cite{xu2018powerful} & Subgraph classification & Key-recovery & \begin{tabular}[c]{@{}l@{}} TLL \\X(N)OR\end{tabular}\\ \hline
\textbf{MuxLink~\cite{muxlink}} & DGCNN~\cite{zhang2018end} & \begin{tabular}[c]{@{}l@{}} Link prediction\\via subgraph classification \end{tabular} & Key-recovery & \begin{tabular}[c]{@{}l@{}} TLL \\MUX\end{tabular} \\ \hline
\textbf{UNTANGLE~\cite{untangle}} & DGCNN~\cite{zhang2018end} & \begin{tabular}[c]{@{}l@{}} Link prediction\\via subgraph classification \end{tabular} & Key-recovery & \begin{tabular}[c]{@{}l@{}}Routing \\obfuscation~\cite{InterLock}\end{tabular} \\ \hline
\end{tabular}%
}
\end{table}

\noindent\textbf{Attack run-time estimation.} The Boolean Satisfiability (SAT)-based attack is one of the seminal attacks on logic locking~\cite{Subramanyan_host_2015}. 
Although the attack can unlock a wide range of logic locking techniques, its attack run-time could vary from seconds to days, depending on the circuit and the logic locking technique. 
To efficiently evaluate the resilience of logic locking techniques against the SAT-based attack, researchers have proposed to use GNNs to predict the run-time of the SAT-based attack for different circuits, locking techniques, and key-sizes. 
The ICNet~\cite{chen2020estimating} platform represents the locked netlist as a graph and encodes each node (\textit{i.e.,} gate) with a key-gate mask (\textit{i.e.,} indicating if the node represents a key-gate) and gate type encoding. 
The GCN performs a regression task, predicting the run-time of the SAT-based attack on the given circuit.
\section{Discussion}
\label{sec:discussion}

In this section, we outline the superiority and shortcomings of GNNs and discuss possible future directions.

\subsection{GNN Superiority}

Traditional ML models, such as convolutional neural networks, can be used to perform the same tasks solved using GNNs. 
However, traditional ML methods require encoding the circuits into a tensor form suitable to be fed to the ML models. 
Thus, the encoding is hand-engineered and follows procrustean rules. 
The core idea of using GNNs is to learn comprehensive circuit encoding and perform the desired tasks in an end-to-end manner. 
In other words, the GNN-based circuit encoding gets optimized for the desired task, unlike the encoding followed by traditional ML methods. 
Thus, the GNN-based hardware security systems outperform traditional baselines~\cite{omla}. 
We provide two examples next.

The authors in~\cite{chen2020estimating} compare the ICNet performance in estimating the SAT attack run-time against several state-of-the-art regression models such as: (i)~linear regression, (ii)~LASSO~\cite{tibshirani1996regression}, (iii)~epsilon-support vector regression~\cite{smola2004tutorial}, (iv)~ridge regression~\cite{ng2004feature}, (v)~elastic net~\cite{zou2005regularization}, (vi)~orthogonal matching pursuit~\cite{mallat1993matching}, (vii)~SGD regression, least angle regression (LARS)~\cite{efron2004least}, and (viii)~Theil-Sen estimators~\cite{dang2008theil}. 
These regression models do not model the circuit as a graph, unlike the ICNet model, which outperforms all these traditional regression methods.

To demonstrate the strength of the GNN model in the context of functional RE over other ML models, the authors in~\cite{gnnre} implemented another supervised
classification method using support vector machines (SVM). 
The authors trained the SVM model using the same feature vectors for nodes as the GNN-RE model. 
Unlike GNN-RE, however, the SVM model allows each node/gate to only reason about its own features and has no access or information to its neighbors' features. 
Still, GNN-RE's feature vector does capture the functionality of gates in the local neighborhood. 
However, the authors demonstrated (via experiments) that such functional information is insufficient to achieve high-accuracy node classification. 
The SVM classifier achieves an accuracy of $81.43\%$, while the GNN's accuracy reaches $98.87\%$.

\subsection{GNN Shortcomings and Solutions}

Although a GNN seems like a natural choice to learn on circuits, it comes with its own challenges, as discussed next.

\begin{enumerate}[leftmargin=*]

\item The 1-Weisfeiler-Lehmann graph isomorphism test bounds the expressive power of GNNs, and thus GNNs can generate identical embeddings for graphs that might be different~\cite{xu2018powerful}. 
Furthermore, platforms such as OMLA~\cite{omla}, MuxLink~\cite{muxlink}, and UNTANGLE~\cite{untangle} propose their own distance encoding methods to enhance the representation capability of the GNNs.

\item Representing a netlist as an undirected graph makes the graphs denser and facilitates message-passing in the network. 
However, we lose the notion of IN/OUT-neighborhood of the original netlist. 
Platforms such as GNN-RE~\cite{gnnre} and GNNUnlock~\cite{gnnunlockp,alrahis2020gnnunlock} encode the input/output degree of the nodes in the extracted feature vectors. 
In addition, platforms such as OMLA~\cite{omla} propose node labeling methods to capture the IN/OUT-neighborhood information in the original netlist that gets lost when represented as an undirected graph. 
Furthermore, platforms such as ABGNN~\cite{he2021graph} represent the circuits as directed graphs. 
However, such platforms train two separate GNNs, where one GNN aggregates information from predecessors and the other from successors.

\item Without encoding the Boolean functionality of the circuits, GNNs fail to recognize optimized circuits with the same functionality but different structure. 
Authors in~\cite{wang2022functionality} proposed a CL-based netlist representation learning framework to extract the functionality and enhance the performance of the GNN.
\end{enumerate}

\subsection{Future Directions}

\textbf{Security of GNNs.} The vulnerability of GNNs themselves (when employed for hardware security) to attacks (e.g., backdoor attacks) has received little to no interest from the research community.
Therefore, it is imperative to ensure that the adoption of GNNs should not introduce additional security vulnerabilities in critical security frameworks.

\noindent\textbf{GNN explainability.} Users can be reluctant to trust GNN-based hardware security systems when used as black-box models. 
It is challenging for users (e.g., circuit designers) to relinquish control to a mysterious GNN model, especially when used in security-critical applications. 
The concepts of \textit{explainable artificial intelligence} have not been explored yet for the GNN-based hardware security systems. 
Employing such explainability methods can aid in improving the practicality of GNN-based platforms. 
Further, explainability can assist in developing robust design-for-trust solutions.
\section{Conclusion}
\label{sec:conclusion}

To the best of our knowledge, our work is the first to compile seminal papers on the applications of graph neural networks (GNNs) in hardware security. 
To that end, we provide a taxonomy that groups GNN-based hardware security systems into four categories: (i)~hardware Trojan (HT) detection systems, (ii)~intellectual property (IP) piracy detection systems, (iii)~reverse engineering (RE), and (iv)~attacks on logic locking. 
We comprehensively review the different platforms within the categorization and summarize the data sets and models used. Finally, we suggest future directions for GNNs in hardware security.
We hope our comprehensive summary will inspire future research for employing GNNs in hardware security, both in developing attacks and defenses.

\bibliographystyle{unsrtnat}
\bibliography{main}

\end{document}